\documentclass[conference]{IEEEtran}
\IEEEoverridecommandlockouts
\usepackage{cite}
\usepackage{amsmath,amssymb,amsfonts}
\usepackage{booktabs}
\usepackage{hyperref}
\usepackage{algorithmic}
\usepackage{graphicx}
\usepackage{textcomp}
\usepackage{xcolor}
\usepackage{multirow,makecell}
\usepackage{amsmath,amsfonts,amssymb,pifont}
\usepackage{multicol}
\makeatletter
\newcommand*{\@rowstyle}{}
\newcommand*{\rowstyle}[1]{
  \gdef\@rowstyle{#1}%
  \@rowstyle\ignorespaces%
}
\newcolumntype{=}{
  >{\gdef\@rowstyle{}}%
}
\newcolumntype{+}{
  >{\@rowstyle}%
}
\makeatother
\def\BibTeX{{\rm B\kern-.05em{\sc i\kern-.025em b}\kern-.08em
    T\kern-.1667em\lower.7ex\hbox{E}\kern-.125emX}}
\newcommand{\cmark}{\text{\ding{51}}}
\newcommand{\xmark}{\text{\ding{55}}}

\begin{document}

\title{MACE: Leveraging Audio for Evaluating Audio Captioning Systems}

\author{
Satvik Dixit, Soham Deshmukh, Bhiksha Raj\\
Carnegie Mellon University \\
\{satvikd, sdeshmuk, bhikshar\}@andrew.cmu.edu
}

\maketitle

\begin{abstract}
The Automated Audio Captioning (AAC) task aims to describe an audio signal using natural language. To evaluate machine-generated captions, the metrics should take into account audio events, acoustic scenes, paralinguistics, signal characteristics, and other audio information. Traditional AAC evaluation relies on natural language generation metrics like ROUGE and BLEU, image captioning metrics such as SPICE and CIDEr, or Sentence-BERT embedding similarity. However, these metrics only compare generated captions to human references, overlooking the audio signal itself. In this work, we propose MACE (\underline{M}ultimodal \underline{A}udio-\underline{C}aption \underline{E}valuation), a novel metric that integrates both audio and reference captions for comprehensive audio caption evaluation. MACE incorporates audio information from audio as well as predicted and reference captions and weights it with a fluency penalty. Our experiments demonstrate MACE's superior performance in predicting human quality judgments compared to traditional metrics. Specifically, MACE achieves a 3.28\% and 4.36\% relative accuracy improvement over the FENSE metric on the AudioCaps-Eval and Clotho-Eval datasets respectively. Moreover, it significantly outperforms all the previous metrics on the audio captioning evaluation task. The metric is opensourced at \href{https://github.com/satvik-dixit/mace}{https://github.com/satvik-dixit/mace}. 
\end{abstract}

\begin{IEEEkeywords}
Automated Audio Captioning, Evaluation Metric, Audio-Language Models
\end{IEEEkeywords}

\vspace{-0.1in}
\section{Introduction} 
Automated Audio Captioning (AAC) task \cite{mei2022automated} is centered on producing natural language descriptions for audio content. 
This process involves identifying audio events \cite{kong2020panns}, acoustic scenes \cite{barchiesi2015acoustic}, temporal relationships \cite{pengi}, etc., within the audio stream. Once trained, an AAC system has numerous applications \cite{mei2022automated}, such as assisting individuals with hearing impairments, enhancing security and surveillance systems, supporting multimedia retrieval, and more.

Building a robust AAC system requires evaluating outputs on three main dimensions: accuracy (covering all audio events, scenes, actions), linguistic quality (grammar, coherence), and readability (clarity and logical flow). Traditional AAC metrics, such as BLEU \cite{bleu}, ROUGE \cite{rouge}, and METEOR \cite{meteor}, emphasize linguistic variation \cite{noaudiocap} through n-gram overlap between candidate and reference sentences. Metrics like SPICE \cite{anderson2016spice} incorporate relational information by parsing captions into a graph containing semantic elements, their attributes, and relations to one another, and evaluates the candidate graph via synonym lemma matching. Recently, FENSE \cite{fense}, developed specifically for audio captioning, leverages sentence-BERT \cite{reimers2019sentence} embeddings to capture semantic similarity between generated and reference captions. Subsequent methods \cite{gontier2023spice+, wang2024x} have built upon this approach, predominantly employing sentence-BERT embeddings. However, a key limitation of existing metrics is their exclusion of audio information and needing reference captions to perform evaluation. 

We hypothesize that incorporating audio information into AAC metrics will enhance semantic accuracy and better align with human judgment. Audio information can be integrated either through direct audio-caption comparison or by grounding text embeddings in audio. For example, a generated caption like ``The crowd is applauding in a stadium” versus the reference ``The crowd is silent in a stadium” would be scored similarly by current metrics like FENSE, despite opposite meanings. Grounding embeddings in audio, however, could yield lower similarity scores for such differences. CLAP \cite{laionclap,msclap22,msclap23, sridhar2024parameter} offers one approach, aligning audio and text in a multimodal space to capture shared audio events and scenes. Yet, solely using CLAP may miss essential linguistic and readability factors \cite{dim}, underscoring the need for metrics that integrate both linguistic quality and audio context for comprehensive evaluation.

In this paper, we propose MACE, a novel and comprehensive metric to evaluate audio captioning system. MACE addresses a fundamental limitation in existing evaluation approaches by incorporating both audio and linguistic information. The metric comprises three components: first, it leverages CLAP audio and text embeddings to assess the relevance of the generated caption with respect to the audio content; second, it employs text embeddings to measure acoustic similarity between the generated and reference captions; third, it applies a fluency error penalty to weight the similarity scores, ensuring grammatical accuracy in the generated captions. We evaluate MACE on two commonly used benchmarks for evaluating AAC metrics - AudioCaps-Eval \cite{fense} and Clotho-Eval \cite{fense} benchmarks. MACE produces SoTA results and outperforms all prior metrics. Specially, MACE achieves a 3.28\% and 4.36\% relative accuracy improvement over the FENSE metric on the AudioCaps-Eval and Clotho-Eval datasets respectively. Moreover, as MACE contains three components, one can evaluate caption quality without requiring reference caption by using two of the three components. 

\section{Related Work}  \vspace{-0.05in}
\noindent \textbf{Linguistic metrics.} Audio captioning evaluation has drawn from fields like NLP and image captioning. Traditional metrics, such as BLEU \cite{bleu} and ROUGE \cite{rouge}, rely on N-gram matching between candidate and reference captions, but struggle in audio captioning, where diverse descriptions can accurately represent the same sound. Advanced metrics like METEOR \cite{meteor} introduced synonym-matching and stemming for better semantic alignment, while CIDEr\cite{cider} used TF-IDF weighting to emphasize rare, meaningful N-grams. SPICE, developed for image captioning, compares “object-graphs” between captions to model conceptual relationships, and SPIDEr\cite{spider}, combining SPICE and CIDEr, aimed to improve robustness and alignment with human judgment.

\noindent \textbf{Embedding-based metrics.} BERT-score \cite{zhang2019bertscore}, BLUERT \cite{sellam2020bleurt} and Sentence-BERT encode candidate and reference sentences as vectors using pre-trained language models, computing distances between these vectors to assess semantic similarity. This approach has shown promise in capturing more subtle semantic relationships that N-gram based methods often miss. In the specific context of audio captioning, FENSE \cite{fense} emerged as a significant advancement. Building upon the embedding-based approach, FENSE incorporates an additional fluency detection mechanism to address the issue of semantically similar but non-fluent captions. Recent efforts have tried to combine parsing-based approaches with embedding methods in two-stage frameworks. SPICE+ \cite{gontier2023spice+} and ACES\cite{wijngaard2023aces} both employ initial parsing stages—either generating parse graphs or extracting explicit sound descriptors—followed by Sentence-BERT embedding comparisons.

\noindent \textbf{LLM-based metrics.} The rapid advancement of large language models (LLMs) like GPT-4 \cite{gpt4} has opened new avenues for evaluation metrics. X-ACE \cite{wang2024x} replaces fixed components in SPICE with LLM-based parsers. CLAIR$_A$ \cite{wu2024clair} represents another innovative approach, directly leveraging an LLM and in-context learning \cite{brown2020language} to produce a numeric score between 0 and 100 quantifying the similarity between candidate and reference captions. This method bypasses intermediate representations, relying instead on the LLM's inherent understanding of language and context. Despite these advancements, a limitation across all these metrics: they operate solely in the textual domain, comparing generated captions to reference captions without considering the audio signal.

\begin{table*}[ht]
\centering
\caption{Benchmarking MACE on Clotho-Eval and AudioCaps -Eval dataset against popular audio caption evaluation metrics}
\label{All-Eval}
\begin{tabular}{l|ccccc|ccccc}
\toprule
& \multicolumn{5}{c|}{Clotho-Eval $\uparrow$} & \multicolumn{5}{c}{AudioCaps-Eval $\uparrow$} \\
Metric & HC & HI & HM & MM & All & HC & HI & HM & MM & All\\
\midrule
BLEU@1 & 51.0 & 90.6 & 65.5 & 50.3 & 59.0 & 58.6 & 90.3 & 77.4 & 50.3 & 62.4 \\
BLEU@4 & 52.9 & 88.9 & 65.1 & 53.2 & 60.5 & 54.7 & 85.8 & 78.7 & 50.6 & 61.6 \\
METEOR & 54.8 & 93.0 & 74.6 & 57.8 & 65.4 & 66.0 & 96.4 & 90.0 & 60.1 & 71.7 \\
ROUGE-L & 56.2 & 90.6 & 69.4 & 50.7 & 60.5 & 61.1 & 91.5 & 82.8 & 52.1 & 64.9 \\
CIDEr & 51.4 & 91.8 & 70.3 & 56.0 & 63.2 & 56.2 & 96.0 & 90.4 & 61.2 & 71.0 \\
SPICE & 44.3 & 84.4 & 65.5 & 48.9 & 56.3 & 50.2 & 83.8 & 77.8 & 49.1 & 59.7 \\
SPICE+ & 46.7 & 88.1 & 70.3 & 48.7 & 57.8 & 59.1 & 85.4 & 83.7 & 49.0 & 62.0 \\
ACES & 56.7 & 95.5 & 82.8 & 69.9 & 74.0 & 64.5 & 95.1 & 89.5 & 82.0 & 83.0 \\
SPIDEr & 53.3 & 93.4 & 70.3 & 57.0 & 64.2 & 56.7 & 93.4 & 70.3 & 57.0 & 64.2 \\
FENSE & 60.5 & 94.7 & 80.2 & 72.8 & 75.7 & 64.5 & 98.4 & 91.6 & 84.6 & 85.3 \\
CLAIR$_{A}$ {\scriptsize(+ Gemini-v1.5)} & 59.0 & 95.9 & 83.2 & 75.1 & 77.4 & 70.4 & 99.2 & 93.7 & 81.5 & 84.9 \\
CLAIR$_{A}$ {\scriptsize(+ GPT-4o)} & 62.4 & 97.1 & \textbf{83.6} & \textbf{77.9} & \textbf{79.7} & 70.9 & \textbf{99.2} & 93.3 & 84.6 & 86.6 \\
\midrule
MACE & \textbf{63.3} & \textbf{98.0} & 80.6 & 77.0 & 79.0 & \textbf{74.4} & \textbf{99.2} & \textbf{94.6} & \textbf{86.3} & \textbf{88.1} \\
\bottomrule
\end{tabular}
\vspace{-0.15in}
\end{table*}

\vspace{-0.1in}
\section{\underline{M}ultimodal \underline{A}udio-\underline{C}aption \underline{E}valuation}
 \vspace{-0.05in}
 The MACE metric utilizes the Contrastive Language-Audio Pre-training (CLAP) model to overcome limitations of text-only evaluation methods by generating embeddings for both audio and text, enabling evaluation that accounts for acoustic content alongside textual descriptions. MACE is computed through three components: (1) CLAP audio and text embeddings assess the relevance of the generated caption to the audio content, (2) text embeddings evaluate acoustic similarity between the generated and reference captions, and (3) a fluency error penalty ensures grammatical accuracy. Additionally, MACE can assess caption quality without a reference by using two of the three components.

\begin{figure}[htbp]
    \centering
    \includegraphics[width=\linewidth]{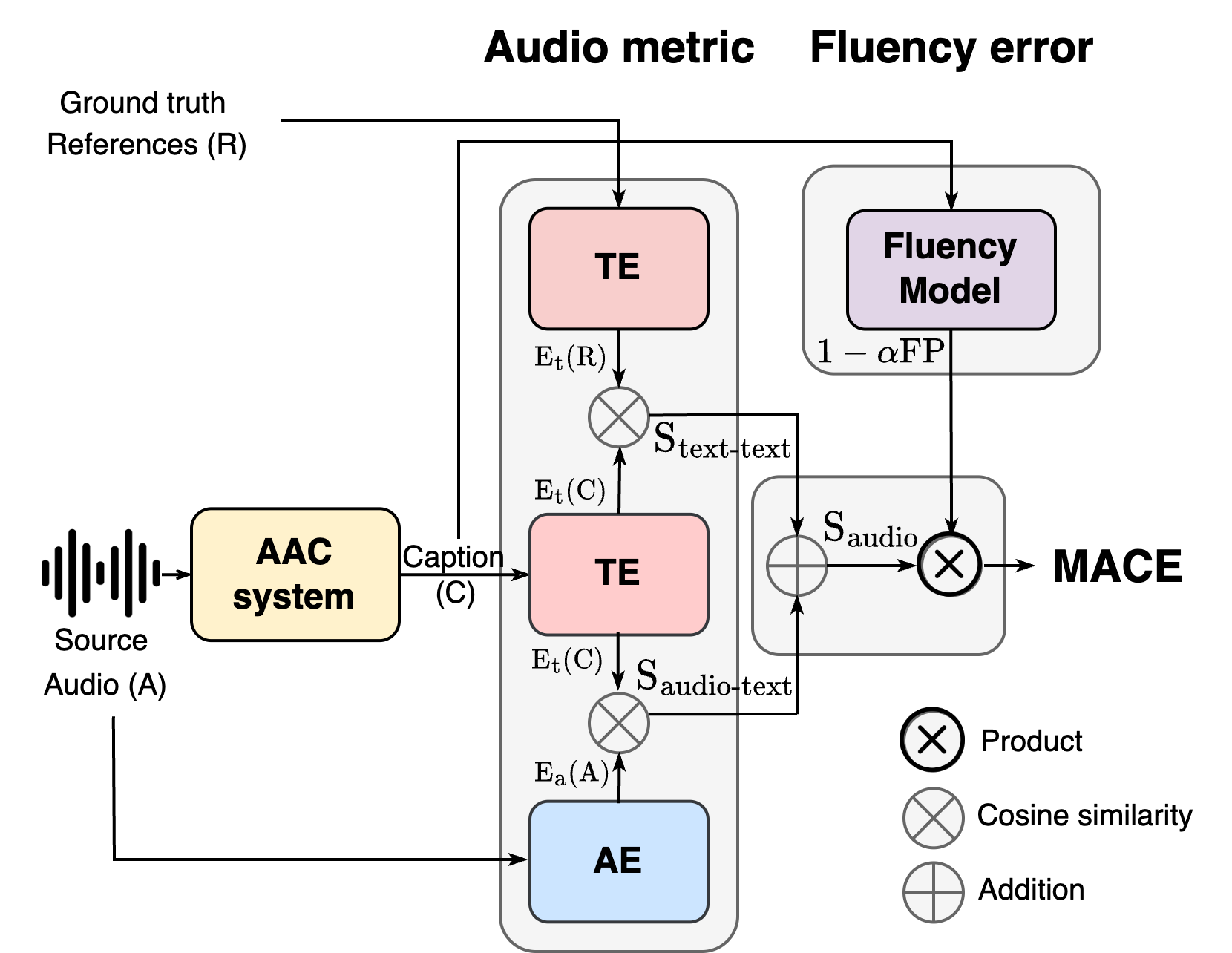} 
    \caption{
    The MACE metric evaluates a generated caption by taking the source audio, ground truth, and generated caption as input. It produces a final score by combining audio-text and text-text metrics with a weighted fluency penalty, integrating both audio content and linguistic quality. Here TE and AE represents the text and audio encoder of CLAP respectively.  \vspace{-0.2in}
    }
    \label{fig:idea} 
\end{figure}

Let $A$ denote the source audio, $C$ the candidate caption, and $R = {R_1, R_2, ..., R_n}$ the set of reference captions. Then MACE metric is computed using three components (Fig. \ref{fig:idea}):
\begin{itemize}
    \item \textbf{Audio-text.} This score measures the similarity between the candidate caption and the audio content. The score is obtained by taking cosine similarity ($\cos(\cdot, \cdot)$) between the CLAP audio embedding ($E_a(A)$) of the source audio and CLAP text embeddings ($E_t(C)$) of the caption:
    \begin{equation}
    \text{S}_\text{audio-text} = \cos(E_a(A), E_t(C)) 
    \end{equation}
    \item \textbf{Text-text.} This score measures the similarity between the candidate caption and the reference caption(s). The score is obtained by taking dot product between the CLAP text embedding ($E_T(C)$) of the candidate caption and the CLAP text embeddings ($E_T(R_i)$) of the reference(s):
    \begin{equation}
    \text{S}_\text{text-text} = \frac{1}{n} \sum_{i=1}^n \cos(E_t(C), E_t(R_i))
    \end{equation}
    \item \textbf{Fluency error.} This score, first introduced in FENSE \cite{fense}, 
    uses a BERT model trained to detect fluency errors like incomplete sentences, repeated elements, and missing conjunctions or verbs, penalizing captions lacking coherence or grammar. If the error probability exceeds a set threshold, the embedding similarity score is reduced by multiplying it by $(1 - \alpha)$, where $\alpha$ is a weighting factor.
\end{itemize}

\noindent The MACE metric is a combination of the above three components. It is computed as:
\begin{equation}
\text{MACE} = \text{S}_\text{audio} \cdot (1 - \alpha \cdot \text{FP})
\end{equation}
where $\alpha$ is a weighting factor, FP is FluencyPenalty which can either be 1 or 0 based on FluencyPenalty scores, and the audio score $\text{S}_\text{audio}$ is average of the audio-text and text-text scores
$$\text{S}_\text{audio} = 0.5\cdot(\text{S}_\text{audio-text} + \text{S}_\text{text-text})$$

MACE captures semantic similarity, acoustic relevance, and linguistic quality within a single metric. By utilizing CLAP's multimodal capabilities, it overcomes the core limitation of existing metrics that overlook the audio signal, offering a more comprehensive and accurate evaluation of the caption quality. 

\section{Experimental setup} \vspace{-0.05in}
We conduct experiments on two widely-used datasets: Clotho-Eval and AudioCaps-Eval \cite{fense}. These datasets provide pairwise human annotations for caption evaluation, comprising 1,671 and 1,750 pairs of audio captions, respectively. To test MACE, we sourced the corresponding audio files for these caption pairs from the original AudioCaps \cite{kim2019audiocaps} and Clotho \cite{drossos2020clotho} datasets. The audio files in AudioCaps-Eval have a duration of 10 seconds, while those in Clotho-Eval range from 15 to 30 seconds. For the CLAP model, we use MSCLAP 2023 \cite{msclap23} model which uses HTSAT \cite{chen2022hts} as audio encoder and GPT2 \cite{radford2019language} as text encoder. The audio is resampled to 44.1 kHz for CLAP. Natively, the CLAP model only supports 7-seconds duration for audio. To support longer audios, we split the audio into 7s clips and take the mean of the corresponding audio embeddings weighted by the duration. 

\begin{figure*}
    \centering
    \begin{minipage}{0.49\textwidth}
        \includegraphics[width=\textwidth]{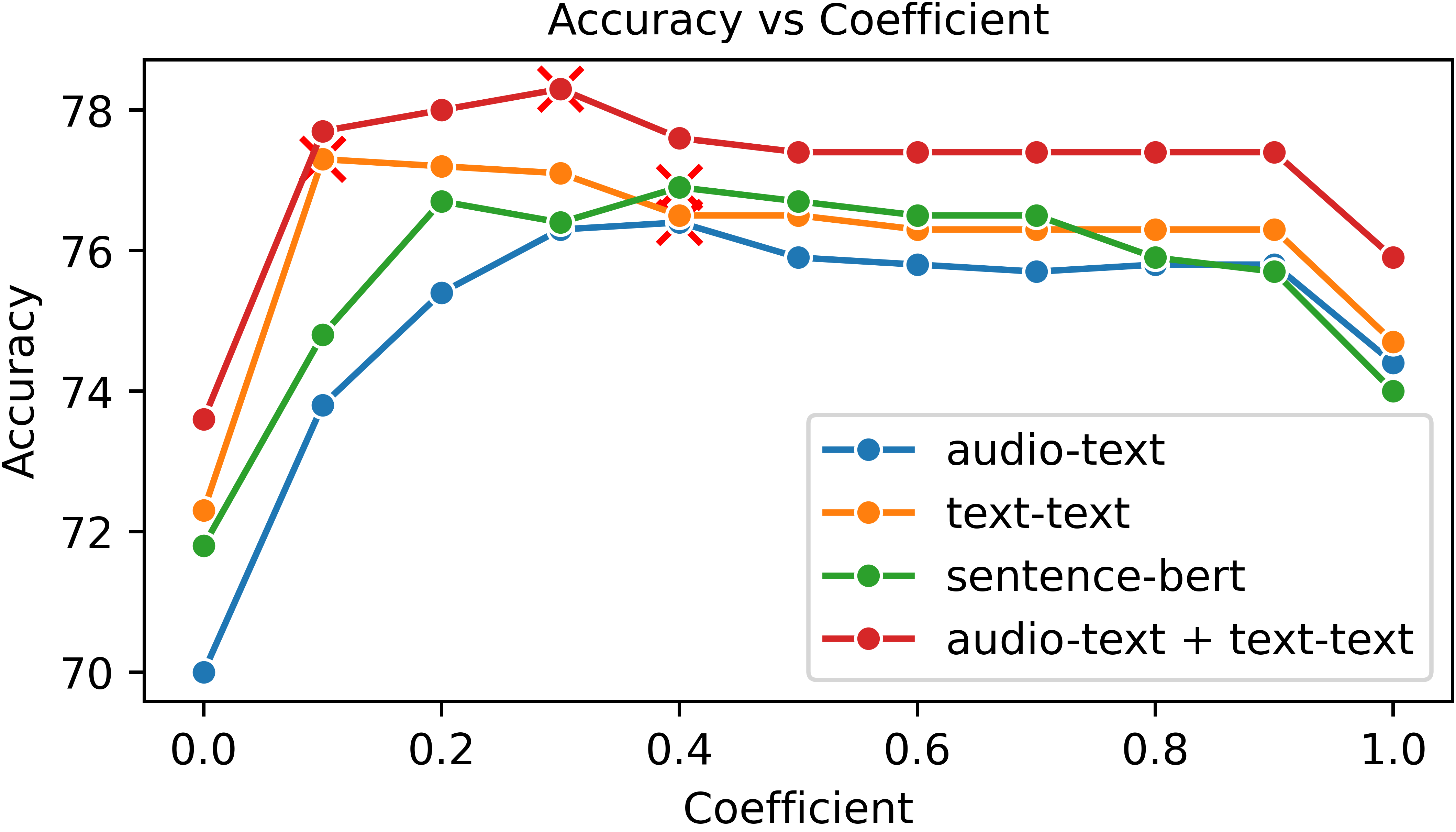}
    \end{minipage}
    \hfill
    \begin{minipage}{0.49\textwidth}
        \includegraphics[width=\textwidth]{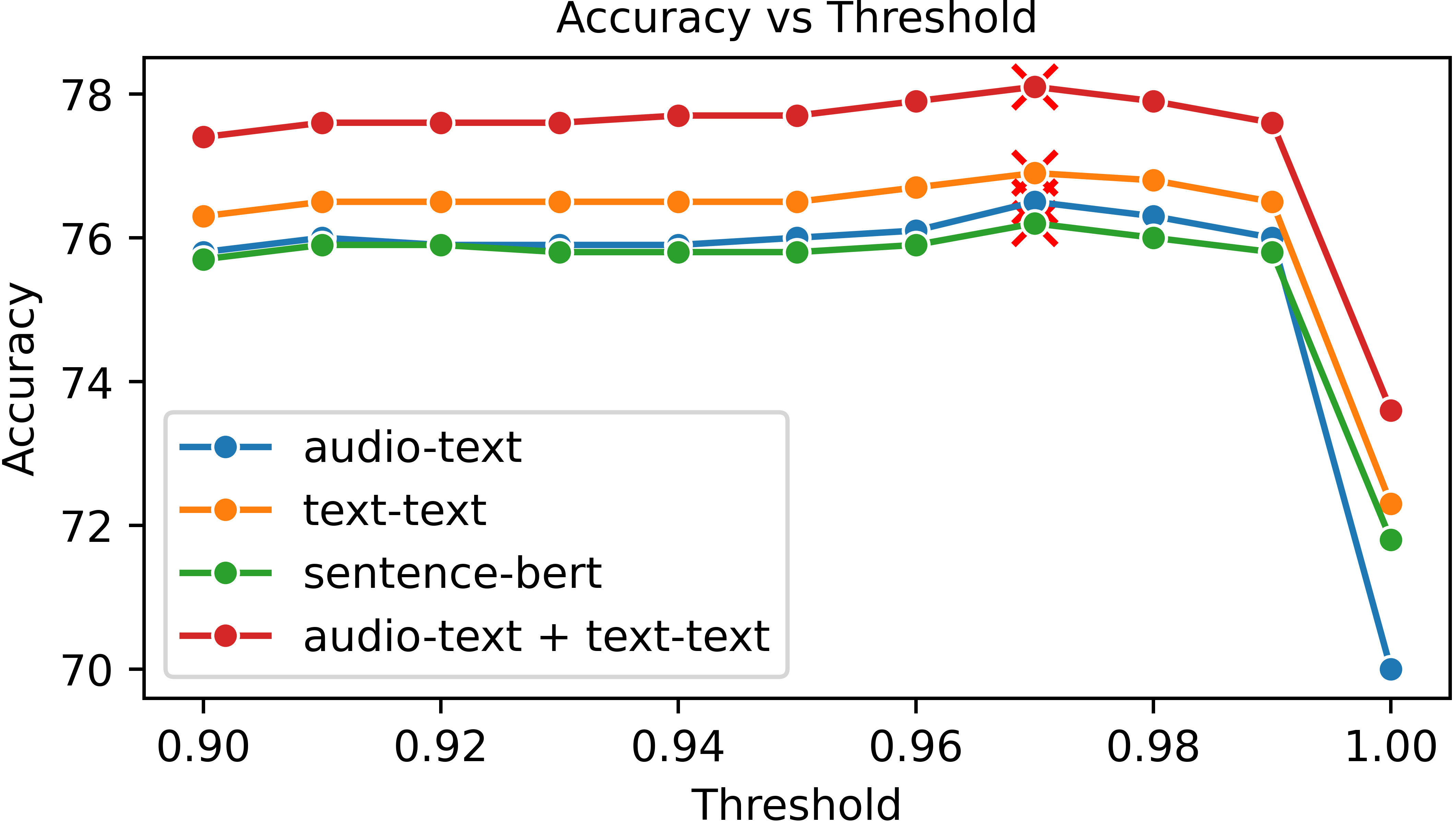}
    \end{minipage}
    \caption{Ablation study on the penalty weight coefficient (left) and threshold (right) parameters of the fluency detection component.}
    \label{fig:thresablation}
    \vspace{-0.1in}
\end{figure*}

\section{Results}\vspace{-0.05in}
\subsection{Evaluating MACE}
To evaluate the effectiveness of MACE, we conducted comprehensive benchmarking experiments on two widely-used datasets: Clotho-Eval and AudioCaps-Eval. The task consists of evaluating a pair of captions and selecting the better caption of the two. One of the captions in each pair is the preferred caption based on the scores assigned by human evaluators. TThe primary metric for comparison is pair accuracy, which measures a metric's ability to correctly identify the human-preferred caption in each pair. To better understand and analyze the results, the tests are categorized into four groups: HC (two correct human captions), HI (one correct and one incorrect human caption), HM (one human and one machine-generated caption), and MM (two machine-generated captions). We compare MACE against a diverse set of metrics such as  BLEU@1, BLEU@4, METEOR, ROUGE-L, CIDEr, SPICE, SPICE+, ACES, SPIDEr, X-ACE, FENSE and CLAIR$_A$ (with Gemini-v 1.5 and GPT-4o).
 all metrics on two categories (HC and HI) and is only slightly behind CLAIRE$_{A}$ on the other two metrics (HM MACEM) and the overall accuracy (79.0\% compared to 79.7\%). The performance of CLAIRE$_{A}$ dMACEs on the ability of GPT-4o to identify which of the two captions in a given pair is better - information which it may have already seen during training. Instead of GPT-4o, if one uses Gemini v1.5 (pro) \cite{gemini}, the accuracy falls down to 77.4\%. 
On AudioCaps-Eval, MACE outperforms other metrics across all categories. The slight drop in performance on Clotho-Eval compared to AudioCaps-Eval, can be attributed to differences in audio content duration (Clotho-Eval contains longer audio clips) and content type (AudioCaps-Eval contains more speech). Notably, MACE demonstrates substantial improvements over the widely used FENSE metric on both the datasets: AudioCaps-Eval (75.7\% to 79.0\%) and Clotho-Eval (85.3\% to 88.1\%).


\subsection{Ablation analysis of MACE's audio components} \label{sec:MACE audio component} \vspace{-0.02in}
In this section, we perform ablation study to understand the contribution of different audio components. The Table \ref{audio_components} shows the performance of MACE$_{AT}$ (using fluency score with S$_{audio-text}$ - the audio-text component of MACE), MACE$_{TT}$ (using fluency score with S$_{text-text}$ - the text-text component of MACE) and the final MACE metric on the Clotho-Eval and AudioCaps-Eval datasets.
Our ablation study reveals a consistent trend in human preference match accuracy across all categories in both datasets: 
$\text{MACE} > \text{MACE}_{TT} > \text{MACE}_{AT}$. The performance gap between $\text{MACE}_{AT}$ and $\text{MACE}_{TT}$ can be attributed to two factors. First, the CLAP audio encoder is trained for 7-second audio inputs, while the dataset contains 10-30 second audio clips, which is not optimal for CLAP and leads to information loss. Second, MACE$_{TT}$ benefits from averaging similarities across five caption-reference pairs, potentially offering more robust evaluations. MACE’s performance over its components indicates that combining audio-text and text-text similarities captures complementary aspects of caption quality, aligning well with human preferences. 


\begin{table}
\centering
\caption{Contribution of Audio components to the MACE score}
\label{audio_components}
\begin{tabular}{=l+l+c+c+c+c+c}
\toprule
Test Data & Metric & HC & HI & HM & MM & All \\
\midrule
\rowstyle{\color{gray}} Clotho & FENSE & 60.5 & 94.7 & 80.2 & 72.8 & 75.7 \\
Clotho & MACE$_{AT}$ & 61.4 & 97.1 & 80.2 & 74.0 & 76.8 \\
Clotho & MACE$_{TT}$ & 60.5 & 97.1 & 79.3 & 75.9 & 77.7 \\
Clotho & MACE & \textbf{63.3} & \textbf{98.0} & \textbf{80.6} & \textbf{77.0} & \textbf{79.0} \\ \midrule
\rowstyle{\color{gray}} AudioCaps & FENSE & 64.5 & 98.4 & 91.6 & 84.6 & 85.3 \\
AudioCaps & MACE$_{AT}$ & 68.5 & 98.4 & 91.2 & 78.7 & 82.6 \\
AudioCaps & MACE$_{TT}$ & 66.5 & \textbf{99.6} & 92.5 & 85.6 & 86.4 \\
AudioCaps & MACE & \textbf{74.4} & 99.2 & \textbf{94.6} & \textbf{86.3} & \textbf{88.1} \\
\bottomrule
\end{tabular}
\vspace{-0.1in}
\end{table}

\subsection{MACE as objective metric}
MACE can also be used as an objective metric \cite{pam} which does not require ground truth or reference caption for computation. To achieve this, we compute MACE but skip computing and adding $\text{S}_\text{text-text}$ score in Fig \ref{fig:idea}. This is equivalent to $\text{MACE}_\text{AT}$ score in Section \ref{sec:MACE audio component} and Table \ref{audio_components}. Compared with FENSE which utilizes reference caption, $\text{MACE}_\text{AT}$ does not utilize reference caption is comparable or only slightly underperforms FENSE. Specifically, $\text{MACE}_\text{AT}$ outperforms FENSE by 1.45\% on Clotho-Eval and underperforms FENSE by 3.16\% on AudioCaps-Eval. This makes MACE ideal for large-scale audio captioning evaluations \cite{mei2024wavcaps, sun2023large,dixit2024vision} where annotations are either unavailable, costly, or restricted due to privacy concerns for users.

\vspace{-0.03in}
\subsection{Ablation analysis of fluency detection} 
To optimize the fluency penalty parameters, we conduct analysis using a 20\% validation set from lotho . We systematically varied the threshold from 0.90 to 1.00 in increments of 0.01, and the penalty coefficient from 0.1 to 1.0 in increments of 0.1. The results of this analysis are presented in Figure  \ref{fig:thresablation}.

Our findings indicate that the optimal threshold for error detection should be set significantly higher (0.97) than the default of 0.9, implying that the BERT model requires greater confidence to accurately identify fluency errors in audio captioning. Additionally, the penalty coefficient should be much lower (0.3) than the default of 0.9. This need for a higher detection threshold and reduced penalty suggests that the BERT model’s predictions do not directly align with human judgments of caption quality for the task of automated audio captioning. This can stem from the nature of AAC, where minor grammatical errors are less impactful on the overall quality of the generated caption. 

\begin{table}
\centering
\caption{Comparison of embeddings for ground-truth vs prediction}
\label{similarities}
\begin{tabular}{lcccc}
\toprule
\textbf{\makecell[l]{Reference: Yelling and \\ then siren and horn}} & H & sBERT & $\text{CL}_{TT}$ & $\text{CL}_{AT}$   \\ \midrule
\makecell[l]{A large engine passes as people \\ speak followed by a siren.} & \cmark & 0.50 & \textbf{0.74} & \textbf{0.36} \\
\makecell[l]{High pitched vibrations and \\ humming of a power tool with \\ some rustling.} &  \xmark & \textbf{0.54} & 0.38 & 0.03 \\
\midrule
\textbf{\makecell[l]{Reference: someone running  \\ across a field made of dirt.}}  & H & sBERT & $\text{CL}_{TT}$ & $\text{CL}_{AT}$\\ \midrule
\makecell[l]{Footsteps crunch on a gravel \\ path at a steady pace.} &  \cmark & 0.34 & \textbf{0.77} & \textbf{0.66} \\
\makecell[l]{A person is walking \\ on a gravel path.} &  \xmark & \textbf{0.51} & 0.75 &  0.58 \\
\bottomrule
\end{tabular}
\vspace{-0.1in}
\end{table}

\vspace{0.1in}
\subsection{Qualitative Evaluation}
One of MACE’s contributions is the integration of audio information (CLAP), rather than relying on traditional text-based embeddings. We perform a qualitative comparison of CLAP embedding similarity with the commonly used Sentence-BERT embedding similarity. Table \ref{similarities} provides representative examples from AudioCaps-Eval and Clotho-Eval, showing a reference caption, a pair of candidate captions, and the similarity scores from Sentence-BERT, $\text{CLAP}_{TT}$ ($\text{CL}_{TT}$) and $\text{CLAP}_{AT}$ ($\text{CL}_{AT}$). The H column indicates human preferred caption out of the two. We see that CLAP$_{AT}$ and CLAP$_{TT}$ have higher similarity values compared to Sentence BERT for the human-preferred caption (marked by \cmark) due to their ability to encode the underlying audio context.

\vspace{-0.05in}
\section{Conclusion} 
\vspace{-0.05in}
We present MACE, a novel metric for evaluating audio captions that integrates both audio and textual information, effectively addressing key limitations in current evaluation methods. By combining audio-text and text-text correspondences with an enhanced fluency penalty, MACE aligns more closely with human perceptions and preferences. Experimental results on standard benchmarks show that MACE outperforms existing metrics in correlation with human judgments, achieving a 3.28\% and 4.36\% relative accuracy improvement over the FENSE metric and significantly surpassing traditional metrics used in the Automated Audio Captioning task.

\newpage

\bibliographystyle{IEEEtran}
{\large\bibliography{refs}}

\end{document}